\documentclass[sigconf]{aamas} 

\usepackage{balance} 
\usepackage{graphicx}
\usepackage{xcolor}
\usepackage{listings}
\usepackage{amsmath} 
\usepackage{graphicx}
\usepackage{geometry}
\usepackage{graphicx, caption, subcaption, booktabs, multirow, float}

\usepackage{amsmath}
\usepackage{algorithm}
\usepackage{algpseudocode}

\usepackage{lineno}
\usepackage{hyperref}

\DeclareMathOperator*{\argmax}{arg\,max}

\title[AAMAS-2025 Formatting Instructions]{Regret-Optimized Portfolio Enhancement through \\Deep Reinforcement Learning and Future Looking Rewards}

\author{Daniil Karzanov}

\affiliation{%
  \institution{AXA Group Operations, EPFL}
  \city{Lausanne}
  \country{Switzerland}}
  \email{daniil.karzanov@axa.com}

\author{Rubén Garzón}

\affiliation{%
  \institution{AXA Group Operations}
  \city{Madrid}
  \country{Spain}}
\email{ruben.garzon@axa.com}

\author{Mikhail Terekhov}

\affiliation{%
  \institution{CLAIRE EPFL}
  \city{Lausanne}
  \country{Switzerland}
}
\email{mikhail.terekhov@epfl.ch}

\author{Caglar Gulcehre}

\affiliation{%
  \institution{CLAIRE EPFL}
  \city{Lausanne}
  \country{Switzerland}
}
\email{caglar.gulcehre@epfl.ch}

\author{Thomas Raffinot}

\affiliation{%
  \institution{ AXA Investment Managers}
  \city{Paris}
  \country{France}
}
\email{thomas.raffinot@axa-im.com}

\author{Marcin Detyniecki}

\affiliation{%
  \institution{ AXA Group Operations}
  \city{Paris}
  \country{France}
}
\email{marcin.detyniecki@axa.com}

\begin{abstract} {

This paper introduces a novel agent-based approach for enhancing existing portfolio strategies using Proximal Policy Optimization (PPO). Rather than focusing solely on traditional portfolio construction, our approach aims to improve an already high-performing strategy through dynamic rebalancing driven by PPO and Oracle agents. Our target is to enhance the traditional 60/40 benchmark (60\% stocks, 40\% bonds) by employing the Regret-based Sharpe reward function. To address the impact of transaction fee frictions and prevent signal loss, we develop a transaction cost scheduler. We introduce a future-looking reward function and employ synthetic data training through a circular block bootstrap method to facilitate the learning of generalizable allocation strategies. We focus on two key evaluation measures: return and maximum drawdown. Given the high stochasticity of financial markets, we train 20 independent agents each period and evaluate their average performance against the benchmark. Our method not only enhances the performance of the existing portfolio strategy through strategic rebalancing but also demonstrates strong results compared to other baselines.

}
\end{abstract}


\newcommand{\BibTeX}{\rm B\kern-.05em{\sc i\kern-.025em b}\kern-.08em\TeX}

\renewcommand\footnotetextcopyrightpermission[1]{}
\setcopyright{none}

\begin{document}


\pagestyle{fancy}
\fancyhead{}


\maketitle 


\section{Introduction}


Strategic asset allocation is a portfolio strategy whereby the investor sets target allocations to various asset classes and rebalances the portfolio periodically. The landscape of finance is continually evolving, driven by the need for more sophisticated and adaptive investment strategies that use machine learning for decision making \cite{de2020machine, karzanov2023headline, Blohm2020ItsAP, hambly2023recent}. Traditional methods of portfolio construction often struggle to keep pace with the dynamic nature of financial markets. This manuscript addresses this challenge by exploring the integration of advanced machine learning techniques to enhance the process of dynamic portfolio construction and rebalancing. By leveraging deep reinforcement learning algorithms (DRL) such as Proximal Policy Optimization (PPO) \cite{schulman2017proximal}  with a future-looking reward function, we aim to demonstrate how RL can optimize financial portfolios, balance risk, and maximize returns in an ever-changing market environment.

The primary objective is to achieve superior returns, while also ensuring that the portfolio's value does not experience significant declines. This secondary objective is measured by the maximum drawdown (MDD). Essentially, our goal is to develop an agent that finds the optimal balance between two inversely related metrics: portfolio return and risk. In this context, we consider MDD to be a more suitable risk measure than portfolio standard deviation, which is commonly used in traditional Markowitz-like approaches \cite{mpt}. MDD is a measure of the maximum observed loss from a peak to a trough in a portfolio's value \cite{chekhlov2005drawdown}, before the portfolio reaches a new peak. It provides an indication of the worst possible loss an investor could have experienced during a specific period. Additionally, we consider the Sharpe Ratio, defined as the ratio of the portfolio's excess return to its standard deviation, to evaluate the risk-adjusted return of the portfolio, allowing for a comprehensive assessment of performance relative to the risk taken. Refer to the \hyperref[sec:glossary]{glossary} in the appendix for details on the Sharpe Ratio and other financial notation and metrics used.

The baseline for comparison is the static allocation of weights based on the conventional 60/40 portfolio (60\% stocks, 40\% bonds). Our overarching aim is to surpass the traditional 60/40 portfolio benchmark in both metrics using a dynamic and adaptive DRL approach.

The novel contributions of this work can be summarized in the following three points: 
\begin{enumerate}
    \item[1.] Introduction of a negative Sharpe regret reward function that leverages Oracle agent's knowledge to encourage optimal allocation and improve out-of-sample performance.
    \item[2.] Integration of real and synthetic data in the training process through a circular block bootstrap method to enhance historical data, thereby accelerating the learning of effective strategies and improving the model’s capacity to extrapolate and generalize beyond observed data trajectories.
    \item[3.] Incorporation of a transaction costs scheduler during training, allowing for the inclusion of frictions that are often overlooked in other studies.
\end{enumerate}

In this paper, we first discuss the relevant literature related to the application of DRL in finance, highlighting key advancements and methodologies. This includes a review of two other benchmark methods used for comparison in our evaluation. We then describe the reinforcement learning framework and the specifics of Proximal Policy Optimization (PPO). In Section 3.2, we delve into the design of our environment, transforming real price data to make it suitable for PPO's learning. This includes the introduction of our novel reward function that incorporates a transaction cost (TC) scheduler, the use of synthetic data generation during training, and other implementation details for the agent. Finally, we provide a discussion of the results, comparing our approach with the benchmark method and two other approaches from the literature, and evaluating different configurations of the agent. We conclude with several noteworthy ideas for future improvements and extensions of this work.

The methodology described in this manuscript can be applied to many other sequential allocation problems, not just financial portfolio construction. Examples include resource allocation in supply chains, task assignment in project management, and bandwidth distribution in telecommunications networks. Each of these scenarios involves distributing limited resources or capacities across various options to achieve optimal outcomes.

\section{Relevant Literature}

Advances in RL and deep RL have greatly impacted the field of AI and ML. Mnih et al. \cite{mnih2015human} achieved human-level control in complex games by leveraging deep Q-networks. Schulman et al. \cite{schulman2017proximal} introduced Proximal Policy Optimization algorithms, addressing stability issues in policy learning. Silver et al. \cite{silver2016mastering} demonstrated beyond human performance in the game of Go through a combination of deep neural networks and tree search.

Numerous studies have explored the application of deep reinforcement learning to both single-asset trading \cite{kochliaridis2023combining, pigorsch2022high, brini2023deep} and portfolio optimization problems \cite{srivastava2020deep, halperin_combining_2022, lu2023evaluation, jiang2017deep}. For example, Benhamou et al. \cite{benhamou2020bridging} employed a policy gradient method to dynamically allocate several systematic strategies. Similarly, Kochliaridis et al. \cite{kochliaridis2023combining} utilized a future-looking reward function that considers future close prices over the next K timesteps to encourage agents to predict future market dynamics in addition to solving the primary optimization problem. Our work incorporates a similar concept by embedding Oracle knowledge (defined as $w^*$ in equation (\ref{eq:regret_weight}) and detailed in Section \ref{sec:reward_function}) into the reward function during the training process. In their Investor-Imitator framework, Ding et al. \cite{ding2018investor} consider an Oracle investor, employing Sharpe and MDD as key evaluation measures.

Another recent study by Sood et al. \cite{sood2023deep} applies a PPO model in a setting similar to ours, utilizing a Differential-Sharpe reward function (\ref{eq:diff_sharpe}) and exponential moving averages, $A_t$, $B_t$ of returns and their standard deviation. The study achieves impressive results, with the agent outperforming traditional mean-variance optimization techniques in both return and drawdown metrics, and also demonstrating a more stable out-of-sample strategy. We adopt the reward function introduced in their paper as one of the benchmarks for comparison in our study.

\begin{equation}
D_t = \frac{\delta S_t}{\delta \eta} = \frac{B_{t-1}\Delta A_t - \frac{1}{2}A_{t-1}\Delta B_t}{(B_{t-1} - A_{t-1}^2)^{3/2}}
\label{eq:diff_sharpe}
\end{equation}
where
\begin{itemize}
    \item $D_t$: Differential Sharpe Ratio at time $t$.
    \item $\delta S_t$: Change in the Sharpe ratio, which is adjusted over time as new information is incorporated.
    \item $\eta$: Incremental time step for daily returns, approximately $\frac{1}{252}$.
    \item $A_t$: Cumulative mean return updated over time, calculated as $A_t = A_{t-1} + \eta \Delta A_t$.
    \item $B_t$: Cumulative second moment (variance) of returns, updated as $B_t = B_{t-1} + \eta \Delta B_t$.
    \item $\Delta A_t$: Change in cumulative mean return, defined as $\Delta A_t = R_t - A_{t-1}$.
    \item $\Delta B_t$: Change in cumulative second moment, defined as $\Delta B_t = R_t^2 - B_{t-1}$.
    \item $R_t$: Return at time $t$.
    \item $A_0$ and $B_0$: Initial values for cumulative mean return and cumulative second moment, both set to 0.
\end{itemize}

Andersson et al. \cite{andersson2023measuring} explore the application of regret theory in financial decision-making, particularly focusing on how regret aversion can influence investor behavior. 

Several studies explore the use of synthetic data in portfolio construction. Pe{\~n}a et el. \cite{pena2024modified} introduce a novel portfolio optimization method using synthetic data generated by a Modified CTGAN algorithm. Similarly, Pagnoncelli et el. \cite{pagnoncelli2023synthetic} demonstrate the advantages of using augmented synthetic data for asset allocation.

Few studies consider the multi-objective case in this niche, which inadequately addresses the problem where investors aim to minimize risk \cite{vcernevivciene2022review}. Bisht et al. \cite{bisht2020deep} and Cornalba et al. \cite{cornalba2024multi} attempt to tackle this issue with multi-objective approaches. Almahdi et al.\cite{almahdi2017adaptive} combine the analysis of both expected maximum drawdown and transaction costs. Similarly, Wu et al. \cite{wu2022embedded} incorporate drawdown into their reward function (\ref{eq:emb_rew}), which prioritizes MDD. If the drawdown level at time $t$, $MDD_t$, exceeds $\alpha$, the desired MDD level, the term in brackets becomes negative, serving as a penalty mechanism.  If the return tends to infinity, the first multiplier tends to the hyperparameter value $k$. Although their study focuses more on trading than on portfolio construction, it offers a practical approach by explicitly including drawdown in the reward function. Consequently, we use their methodology as one of the baselines for our study, using a dynamic $\alpha$ equal to the MDD of the 60/40 benchmark.

\begin{equation}
\text{\textbf{Reward}}_t = \frac{k}{1 + e^{-r}} (-e^{MDD_t} + e^{\alpha})
\label{eq:emb_rew}
\end{equation}

Except for a few studies \cite{lucarelli2020deep, lucarelli2019deep, jiang2017deep}, most papers neglect transaction costs and other similar frictions in their analysis. In contrast, we introduce a transaction cost (TC) term and devise an elegant method to integrate it into the learning process without attenuating the contribution of the main reward during exploration (e.g. instantaneous return or Sharpe without TC term).

\begin{figure*}[ht]
    \centering

    \includegraphics[width=\textwidth]{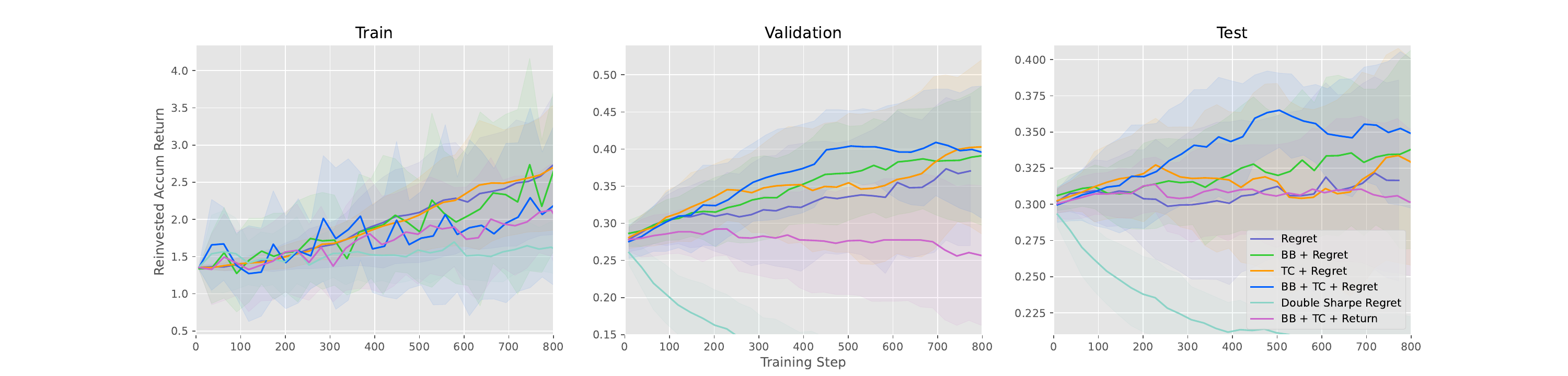}
    \caption{The evolution of accumulated (financial) return over training. Ablation study: removal of specific model components to assess their impact. TC: inclusion of transaction cost schedule. BB: block bootstrap synthetic data. Regret: our reward function. Return: purely optimizing for returns, without including MDD or risk in the reward. Each configuration is averaged over 20 runs. \\
    The full configuration (TC + BB + Regret) generalizes better despite underperformance during training due to increased variability. Synthetic data functions as a form of regularization, preventing memorization of non-reproducible strategies. When used alone, BB and TC are less effective than when combined. However, the combination of TC, BB, and the Return reward function (which specifically optimizes the value displayed on the Y-axis of the plot) tends to overfit on train, resulting in an inability to generate profitable out-of-sample strategies.}
    \label{fig:novelt}
    \Description{The evolution of accumulated (financial) return over training. Ablation study: removal of specific model components to assess their impact. TC: inclusion of transaction cost schedule. BB: block bootstrap synthetic data. Regret: our reward function. Return: purely optimizing for returns, without including MDD or risk in the reward. Each configuration is averaged over 20 runs. \\
    The full configuration (TC + BB + Regret) generalizes better despite underperformance during training due to increased variability. Synthetic data functions as a form of regularization, preventing memorization of non-reproducible strategies. When used alone, BB and TC are less effective than when combined. However, the combination of TC, BB, and the Return reward function (which specifically optimizes the value displayed on the Y-axis of the plot) tends to overfit on train, resulting in an inability to generate profitable out-of-sample strategies.}
\end{figure*}

\section{Background}
\subsection{Portfolio Optimization}
Portfolio optimization plays a crucial role in financial management, focusing on the allocation of assets to maximize returns while effectively managing risk. Mathematical portfolio optimization seeks to allocate assets in a way that maximizes expected return while minimizing risk. Risk can be represented by various measures, such as portfolio standard deviation or MDD, both of which are typically inversely related to expected return. The classic formulation is based on mean-variance single-period optimization, introduced by Markowitz \cite{mpt}. The goal is to solve the trade-off between portfolio risk $w' \Sigma w$ and return $w' \mu$:

\begin{equation}
\min_{w} \frac{1}{2} w' \Sigma w - \lambda w' \mu
\end{equation}

where \(w\) represents a vector of asset weights, denoting the proportion of the portfolio invested in each asset, \(\Sigma\) is the covariance matrix of asset returns, \(\mu\) is the vector of expected returns, and \(\lambda\) is a risk-aversion parameter. Constraints such as \(\sum w_i = 1\) (full investment) and \(w_i \geq 0\) (no short-selling) are typically included.
Classical portfolio optimization approaches rely solely on historical data and do not incorporate forecasting, unlike neural networks. We aim to apply reinforcement learning to address the multi-period portfolio optimization problem, which involves making investment decisions over several time periods (e.g. daily) to adapt to and predict changing market conditions. This approach allows for dynamically rebalancing of the portfolio based on current asset information ( \(\mu\) and \(\Sigma\)) and relevant exogenous predictors (discussed in section \ref{ch:obs_space}).

\subsection{Reinforcement Learning}

Reinforcement learning is a subfield of machine learning where an agent learns to make decisions by interacting with an environment to maximize cumulative rewards \cite{sutton2018reinforcement}. The agent operates by taking actions $w_t$\footnote{In traditional reinforcement learning terminology, $w_t$ is referred to as $a_t$.} at each time step $t$, receiving a state $s_t$ from the environment, and obtaining a reward $r_t$. The goal of the agent is to learn a policy $\pi$ that maximizes the expected cumulative reward, defined as the return $G_t$, over time.

The return $G_t$ is the total discounted reward from time step $t$ onwards and is given by:
\begin{equation}
G_t = \sum_{k=0}^{\infty} \gamma^k r_{t+k+1},
\end{equation}
where $\gamma$ (0 $\leq \gamma < 1$) is the discount factor, which determines the present value of future rewards. The expected return, or the value function $V(s)$, under a policy $\pi$ is defined as:
\begin{equation}
V^\pi(s) = \mathbb{E}_\pi [G_t | s_t = s].
\end{equation}
The objective in reinforcement learning is to find an optimal policy $\pi^*$ that maximizes the value function $V(s)$ for all states $s$.

In the context of dynamic portfolio construction, the reinforcement learning agent's task is to dynamically adjust the portfolio weights based on the observed market states $s_t$ to maximize the cumulative reward $G_t$. The exact form of the reward function $r_t$ may vary and its design may result in different agent behaviors.

\subsection{Proximal Policy Optimization }
Proximal Policy Optimization (PPO) is one of the most effective and widely used algorithms in reinforcement learning. It was introduced as a method to improve both the stability and performance of policy gradient methods. PPO achieves this by using a surrogate objective function, which helps in balancing the trade-off between exploration and exploitation. The core idea behind PPO is to ensure that the new policy does not diverge too much from the old policy during training. This is accomplished through a mechanism known as clipping. Specifically, PPO optimizes the following objective:
\begin{equation}
    \mathcal{L}^{CLIP}(\theta) = \mathbb{E}_t \left[ \min \left( r_t(\theta) \hat{A}_t, \text{clip}(r_t(\theta), 1 - \epsilon, 1 + \epsilon) \hat{A}_t \right) \right].
\end{equation}
In this objective, $r_t(\theta) = \frac{\pi_\theta(w_t | s_t)}{\pi_{\theta_{\text{old}}}(w_t | s_t)}$ represents the probability ratio between the new policy $\pi_\theta$ and the old policy $\pi_{\theta_{\text{old}}}$. The term $\hat{A}_t$ is the advantage estimate, which indicates how much better the current action is compared to the expected action. The hyperparameter $\epsilon$ controls the clipping range, thereby ensuring that the updates to the policy are conservative and stable. The clipping mechanism is crucial as it prevents excessively large updates that can destabilize the learning process. By limiting the change to a specified range, PPO maintains a balance between optimizing the policy and keeping the updates within a reasonable bound.

PPO has been shown to be robust and effective in a variety of complex environments, making it a popular choice for tasks that require dynamic decision-making and adaptation, such as dynamic portfolio construction. Its ability to maintain stability while still performing well in challenging scenarios is a key reason for its widespread adoption in the reinforcement learning community.

\begin{table*}[ht]
\caption{Performance comparison of the 60/40 benchmark and the PPO agent with differential Sharpe  eq. (\ref{eq:diff_sharpe}), embedded drawdown eq. (\ref{eq:emb_rew}) and negative Sharpe regret (ours) reward functions. 
}
\centering
\footnotesize

\begin{tabular}{lllllllllll}
\toprule
 &  & \multicolumn{3}{c}{train} & \multicolumn{3}{c}{valid} & \multicolumn{3}{c}{test} \\
  &  & Phase 1 & Phase 2 & Phase 3 & Phase 1 & Phase 2 & Phase 3 & Phase 1 & Phase 2 & Phase 3 \\
  &  & \scriptsize{(pre-pandemic)} & \scriptsize{(pandemic)} & \scriptsize{(post-pandemic)} & \scriptsize{(pre-pandemic)} & \scriptsize{(pandemic)} & \scriptsize{(post-pandemic)} & \scriptsize{(pre-pandemic)} & \scriptsize{(pandemic)} & \scriptsize{(post-pandemic) } \\
\midrule

\multirow[t]{4}{*}{Annual return} & 60/40 & 0.052 & 0.054 & 0.081 & 0.096 & 0.076 & 0.077 & 0.056 & 0.105 & -0.026 \\
 & Diff. Sharpe & \textbf{0.054} & 0.054 & 0.076 & 0.083 & 0.066 & 0.077 & 0.053 & 0.093 & \textbf{-0.024} \\
 & Emb. DD & \textbf{0.056} & 0.051 & 0.057 & 0.042 & 0.062 & 0.055 & 0.041 & 0.072 & -0.038 \\
 & Regret & \textbf{0.071} & \textbf{0.064} & \textbf{0.091} & \textbf{0.128} & 0.069 & \textbf{0.093} & \textbf{0.064} & \textbf{0.128} & \textbf{-0.007} \\
\cline{1-11}

\multirow[t]{4}{*}{Sharpe ratio} & 60/40 & 0.541 & 0.568 & 0.938 & 1.447 & 1.181 & 0.776 & 0.803 & 0.868 & -0.183 \\
 & Diff. Sharpe & \textbf{0.603} & \textbf{0.658} & \textbf{0.97} & 1.405 & \textbf{1.21} & \textbf{0.833} & \textbf{0.869} & \textbf{0.876} & \textbf{-0.179} \\
 & Emb. DD & \textbf{0.849} & \textbf{0.742} & 0.936 & 1.005 & \textbf{1.362} & \textbf{0.847} & \textbf{0.923} & 0.832 & -0.368 \\
 & Regret & \textbf{0.614} & \textbf{0.593} & 0.882 & \textbf{1.558} & 0.841 & 0.756 & \textbf{0.857} & 0.844 & \textbf{-0.005} \\
\cline{1-11}

\multirow[t]{4}{*}{Calmar ratio} & 60/40 & 0.136 & 0.142 & 0.475 & 1.399 & 0.625 & 0.355 & 0.461 & 0.486 & -0.114 \\
 & Diff. Sharpe &  \textbf{0.159} &  \textbf{0.180} &0.442 &  \textbf{1.453} &  \textbf{0.631} &  \textbf{0.413} &  \textbf{0.512} &  \textbf{0.526} &  \textbf{-0.109} \\
 & Emb. DD &  \textbf{0.257} &  \textbf{0.205} & 0.44 & 0.983 &  \textbf{0.756} &  \textbf{0.454} &  \textbf{0.560} &  \textbf{0.535} & -0.17 \\
 & Regret &  \textbf{0.192} &  \textbf{0.185} & 0.429 &  \textbf{1.602} & 0.451 &  \textbf{0.364} &  \textbf{0.607} & 0.483 &  \textbf{-0.035} \\
\cline{1-11}


\multirow[t]{4}{*}{Max drawdown} & 60/40 & -0.384 & -0.382 & -0.171 & -0.069 & -0.121 & -0.216 & -0.122 & -0.216 & -0.225 \\
 & Diff. Sharpe & \textbf{-0.338} & \textbf{-0.303} & -0.173 & \textbf{-0.057} & \textbf{-0.105} & \textbf{-0.187} & \textbf{-0.104} & \textbf{-0.178} & \textbf{-0.215} \\
 & Emb. DD & \textbf{-0.225} & \textbf{-0.256} & \textbf{-0.129} & \textbf{-0.043} & \textbf{-0.083} & \textbf{-0.122} & \textbf{-0.073} & \textbf{-0.134} & \textbf{-0.224} \\
 & Regret & -0.387 & \textbf{-0.361} & -0.211 & -0.080 & -0.152 & -0.255 & \textbf{-0.108} & -0.266 & \textbf{-0.205} \\
\cline{1-11}

\multirow[t]{4}{*}{Omega ratio} & 60/40 & 1.156 & 1.158 & 1.266 & 1.418 & 1.331 & 1.240 & 1.220 & 1.286 & 0.957 \\
 & Diff. Sharpe & \textbf{1.174} & \textbf{1.183} & \textbf{1.279} & 1.409 & \textbf{1.334} & \textbf{1.264} & \textbf{1.237} & \textbf{1.288} & \textbf{0.958} \\
 & Emb. DD & \textbf{1.237} & \textbf{1.204} & 1.259 & 1.281 & \textbf{1.374} & \textbf{1.265} & \textbf{1.245} & 1.271 & 0.916 \\
 & Regret & \textbf{1.198} & \textbf{1.173} & 1.260 & \textbf{1.488} & 1.238 & 1.235 & \textbf{1.246} & 1.277 & \textbf{0.999} \\
\cline{1-11}


\bottomrule
\end{tabular}

\label{tab:results1}
\end{table*}



\section{Methodology}
One of the most crucial aspects of reinforcement learning applications is the definition of the environment. Its design and hyperparameters significantly influence the agent's performance.

\subsection{Action Space}
We consider prices of $K=3$ trading strategies:  Only \textit{Developed Markets Equity} when the agent is more bullish\footnote{expecting a rise in prices, and thus choosing a more volatile asset}, the \textit{60/40 Portfolio} (i.e. the portfolio that we are trying to improve), and Only \textit{Global Government Bonds} (Govies) as a low-risk asset to potentially avoid sharp drops in portfolio value.
Shorting \footnote{ taking negative weight $w^i$ by borrowing asset $i$}  is not allowed, hence our action space $ \mathcal{A}$ is represented by a non-negative continuous vector $w \in \{w \in \mathbb{R}_{+}^K : \sum_{i=1}^{K} w^i = 1\}$. 
While we consider a standard problem of allocation between risky, balanced and conservative assets, the approach can be scaled to more high-dimensional allocation.

\subsection{Observation Space}
\label{ch:obs_space}
 In addition to asset information, we include three classical contextual indexes that, while not part of the portfolio, may provide signals for asset allocation: \textit{High-Yield Bond Spread} \cite{gilchrist2012credit}, \textit{VIX volatility index} \cite{whaley2000investor}, and \textit{Merrill Lynch Option Volatility Estimate} \cite{driessen2009price}.
 

We transform the price data for both asset and index tickers into daily strategy returns, $\mu_t \in \mathbb{R}^{3}$ and $\alpha_t \in \mathbb{R}^{3}$ respectively, as well as the return standard deviations over the last 60 days for assets $\bar \sigma^{t-60}_{t} \in \mathbb{R}^{3}_+$ and for indexes ${\bar q^{t-60}}_{t} \in \mathbb{R}^{3}_+$. Additionally, we calculate the rolling average asset returns over the past 40 days, $\bar \mu^{t-40}_t \in \mathbb{R}^{3}$, to smooth out noise in the daily returns. The agent also considers the previous allocation $w_{t-1} \in \{w \in \mathbb{R}_{+}^3 : \sum_{i=1}^{3} w^i = 1\}$ and the current transaction costs, $TC_{train}(t) \in \mathbb{R}_+$, to decide whether adjusting the current allocation is warranted or if it remains approximately optimal. The agent iterates through the transformed historical dataset, rebalancing the portfolio based on current observations, and loops back to the beginning upon reaching the end, treating the data as ongoing observations. At each timestep, the agent observes a vector $o_t = [\mu_t, \ \alpha_t, \ \bar \mu^{t-40}_t, \ \bar \sigma^{t-60}_{t}, \ {\bar q^{t-60}}_{t}, \ w_{t-1}, \ TC_{train}(t)] \in \mathbb{R}^{19}$.

Daily portfolio rebalancing can be excessively sensitive to short-term noise and costly due to fees. To address this, we employed data aggregation; however, this approach reduces the number of training points by the frequency of trading days we choose to use. We settled on a bi-daily (2 bd) interval because aggregating data over two-day intervals smooths the data while avoiding an excessive reduction in training points.

\subsection{TC Schedule}
In our first experiments, we observed that including transaction costs (TC) at the start of training negatively impacted the learning process and diminished the signal from the main reward term (e.g., return or Sharpe ratio), particularly during the model's exploration stage. Fine-tuning the model with full transaction costs from the start proved challenging, as it often led to either constant allocations or poor generalization (see Regret and BB + Regret in the validation/test plots, Fig. \ref{fig:novelt}). Inspired by the principles of Curriculum Learning \cite{bengio2009curriculum, koenecke2020curriculum}, where the complexity of tasks gradually increases from simple to more real-world scenarios, we introduce a transaction cost scheduler that incrementally raises the training transaction cost at each step according to 
\begin{equation}
TC_{\text{train}}(x) = 
\frac{TC_{\text{max}}}{S^{a}} \cdot x^{a} \quad \text{if } 0 \leq x \leq S.
\label{eq:tc_formula}
\end{equation}
The costs are increased until a ramp limit given by $S = 100 \cdot \text{episode\_length}$ is reached. After the ramp limit, the maximal costs, $TC_{\text{max}} = TC_{\text{eval}} = 0.0025$ (a fair value for traditional brokers and large institutional traders), are applied.

\subsection{Reward Function}
\label{sec:reward_function}
A key distinction between our problem and many other RL applications is the ability to know the reward of any action, not just the one taken by the agent. Real market participants often reflect on the optimal allocation they could have chosen yesterday based on today's returns. Building on this concept, we propose a negative Sharpe-based regret reward function (\ref{eq:regret}) conditioned on the previous timestep's action. 

\begin{equation}
\text{\textbf{Reward}}_t = -\bar \mu_t^{t+n}(w^* - w_t)'
\label{eq:regret}
\end{equation}
where
\begin{equation}
\begin{aligned}
w^* &= \argmax_{w} \ \mathbf{Sharpe}(w, \bar \mu_t^{t+n}, \bar \Sigma_{t-3n}^{t+3n}) - TC_{\text{train}}(t) \cdot \|w - w_{t-1}\|_1 \\
 &= \argmax_{w}  \frac{w' \bar \mu_t^{t+n} - R_f}{\sqrt{w' \bar \Sigma_{t-3n}^{t+3n} w}} - TC_{\text{train}}(t) \cdot \|w - w_{t-1}\|_1
\end{aligned}
\label{eq:regret_weight}
\end{equation}

The reward is calculated as the difference between the average returns of the optimal allocation and the agent's allocation over the next $n$ days, assuming both allocations are fixed from today. We use the Sharpe ratio because it is one of the most popular and efficient measures that balance return and risk. To incentivize the agent to optimize the portfolio for future relevance, we use a forward-looking return vector $\bar{\mu}_t^{t+n}$ and a covariance matrix $\bar{\Sigma}_{t-3n}^{t+3n}$ in the Sharpe ratio calculation, rather than the typical current or simple return. While we consider only the forward-looking average return over the next $n=14$ business days, we use a broader interval $(t - 3n, \ t + 3n)$ for a more precise estimate of the covariance matrix. Additionally, to account for transaction costs, we include a regularization term in the optimal action's expression (\ref{eq:regret_weight}),  ensuring that the next Oracle-optimal allocation $w^*$ can be achieved despite the fees required to adjust the previous allocation, with the transaction cost term proportional to the difference in weights between two consecutive allocations to penalize large adjustments.

Since we do not have access to future return information during testing, we set the reward to zero in the environment when deploying the trained model. While future data can be integrated during training, we avoid using this data at inference time to prevent leakage between training, validation, and testing. We ensure no overlap among these datasets, which means that toward the end of training, we have fewer points to look into the future. Thus, we use all available training points without incorporating data that cannot technically be used. Consequently, we evaluate the pipeline using various financial ratios instead of total episodic reward.

\subsection{Training on Synthetic Data}
A significant conceptual challenge in applying reinforcement learning to this problem is the limited size of available data. Unlike other RL environments in fields such as robotics and games, which offer variability and diverse paths during training as observations are influenced by the agent's actions, our environment lacks price impact and remains unaffected by allocations. As a result, the agent iterates through the same data repeatedly, encountering nearly identical trajectories in each episode. This setup significantly increases the risk of overfitting to the training dynamics rather than learning robust and generalizable trading strategies.

To address this issue, we propose training on synthetic data that closely mimics the underlying real data distribution. Initially, we considered using the Gaussian copula \cite{rey2015copula}; however, this method was quickly dismissed as it fails to accurately represent the distribution in the tails, which are critical during crisis events that offer opportunities for abnormal returns. Instead, we opted for a circular block bootstrap \cite{arch} of the underlying training data, applied every 10 episodes. We experimented with various block sizes and found that larger blocks,  70-90\% of the original training set, led to improved agent performance as they better preserve long-range temporal dependencies of the training set.

The training process involves an iterative approach to improve the agent's performance. Initially, the agent is trained on real data for 10 episodes. Subsequently, the training shifts to synthetic data for another 10 episodes. After this cycle, the synthetic data is regenerated to introduce new variability, and the agent undergoes another 10 episodes of training on the new synthetic data. This process of regenerating synthetic data and training continues until the required number of training episodes is reached. Upon completion, the agent transitions to the next phase, starting the cycle anew with training on real data.

\begin{figure}[ht]
  \centering
  \begin{subfigure}{\linewidth}
    \centering
  \includegraphics[width=\linewidth]{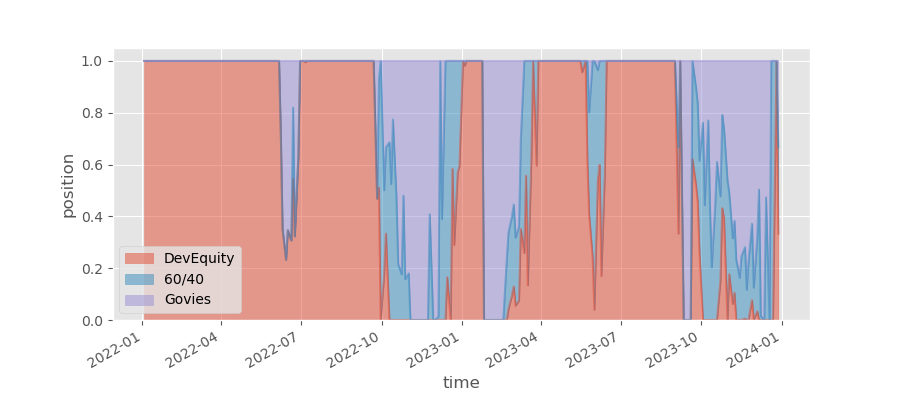}
  \caption{Oracle: $w^*$ defined as (\ref{eq:regret_weight})}
  \label{fig:Worcale}
  \end{subfigure}
  \begin{subfigure}{\linewidth}
    \centering
 \includegraphics[width=\linewidth]{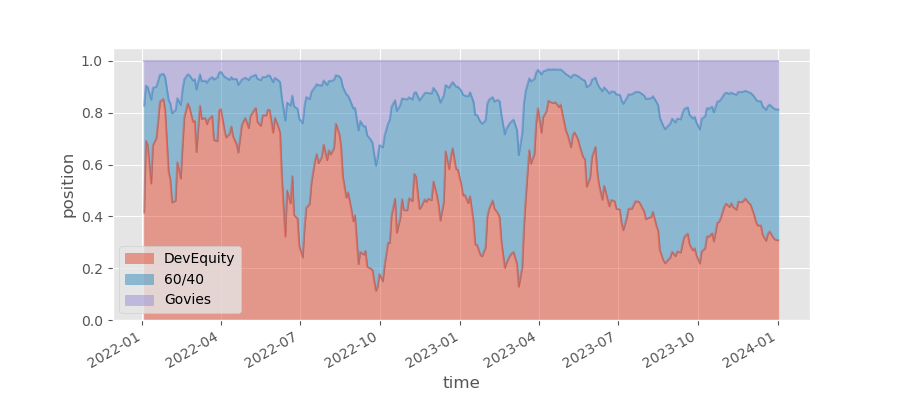}
  \caption{PPO with negative Sharpe Regret reward function}
  \label{fig:Alloc}
  \end{subfigure}

  \caption{Example of allocation during the testing period of phase 3. The regret-based agent demonstrates a general alignment with the optimal allocation but adopts a less aggressive stance. During bullish market periods, it increases its positions in risky assets, capitalizing on favorable conditions. Conversely, when market uncertainty arises, the agent shifts to more conservative allocations, showcasing its adaptability to changing market dynamics.}
  \Description{Example of allocation during the testing period of phase 3. The regret-based agent demonstrates a general alignment with the optimal allocation but adopts a less aggressive stance. During bullish market periods, it increases its positions in risky assets, capitalizing on favorable conditions. Conversely, when market uncertainty arises, the agent shifts to more conservative allocations, showcasing its adaptability to changing market dynamics.}
  \label{fig:two-allocs}
  
\end{figure}

\subsection{Implementation Details} 
The complete training loop is detailed in Algorithm \ref{alg1}. We employ fully connected feedforward neural networks for both the policy and critic architectures. These networks are designed with multiple hidden layers to effectively extract features from the state space, capturing the intricate patterns and dynamics of the environment. The hidden layers use the Tanh activation function, which maps the input values to a range between -1 and 1, aiding in normalization and maintaining stable gradient flow during backpropagation. This design ensures that the networks can learn robust feature representations, contributing to the overall stability and performance of the PPO algorithm in our experiments.

We employ a sliding window approach with a train-validation-test split in our pipeline, splitting the available data into phases as detailed in Table \ref{tab:timing-phases}. This method mitigates the bias towards older, potentially less relevant observations. The phased split enables a robust evaluation of all approaches by testing the agent across three distinct and representative market scenarios. Phase 1 represents a period of asset growth under favorable market conditions. Phase 2 corresponds to the COVID-19 crisis, characterized by sharp declines in asset values. Finally, Phase 3 reflects a market downturn in which all assets experienced price drops, emphasizing the agent's ability to minimize losses in adverse conditions. We transfer the weights of the value and policy networks after Phases 1 and 2. The best model, evaluated on the validation set based on the return-risk trade-off, is used as a pretrain for the subsequent phase. This approach prevents the need to learn the model of the world from scratch, thereby reducing the number of training episodes. We opted for the sliding window approach over the expanding window because the latter approach often leads to overfitting to the early dynamics by exposing the agent to the earliest observations more frequently. Consequently, we exclude data from the 90s in the later phases but retain the knowledge by transferring the policy and value networks' weights to the subsequent phases. Additionally, to promote exploration when transitioning to another phase, we implement an entropy regularization schedule. This schedule sets a non-zero entropy coefficient, $\beta_{\text{entropy}}$, in the PPO total loss function (\ref{eq:ppo_loss}) and then linearly decreases it to zero until 10\% of the number of training episodes. We use a Critic loss (\ref{eq:critic_loss}) and a bonus based on the $ \mathcal{L}_{\text{entropy}} = - H(\pi(w|s))$ of the policy distribution. 

\begin{equation}
\mathcal{L} = \mathcal{L}_{\text{policy}} + \beta_{\text{entropy}} \cdot \mathcal{L}_{\text{entropy}} + \beta_{\text{value}} \cdot \mathcal{L}_{\text{value}}
\label{eq:ppo_loss}
\end{equation}

\begin{equation}
\mathcal{L^{\text{value}}} = \mathbb{E}_{t} \left[ \left( V_{\theta}(s_t) - V^{\text{target}}_t \right)^2 \right]
\label{eq:critic_loss}
\end{equation}

Normalizing the advantages in PPO enhances the results on the validation set. We adjust the transaction cost schedule to be more concave in Phases 2 and 3, as the agent is not learning from scratch. Furthermore, we discovered that when selecting the model from the Pareto front for the next phase, it is preferable to choose a less risk-averse model, as more conservative models tend to become stuck in suboptimal allocations and encounter difficulties in training during the new phase.

\begin{algorithm}
\caption{Agent Training Pseudo-Code}
\label{alg1}
\begin{algorithmic}[1]
\State Given original historical dataset $\mathcal{D}_{\text{orig}}$

\State Initialize network parameters: $\theta$ for policy network, $\phi$ for value network
\State Initialize maximum training iterations $M$ and training batch $D$

\State Set $\tilde{\mathcal{D}} = \mathcal{D}_{\text{orig}}$
\For{training iteration $= 1$ to $M$}
    \State Clear training batch $D$
    \For{each RL collect step $t$}
        \If{$t \ \%  \ \text{episode\_length} \times 10 == 0$}
            \If{$Bernoulli(0.7) == 1$}
                \State $\tilde{\mathcal{D}} = BlockBootstrap(\mathcal{D}_{\text{orig}})$
            \Else
                \State $\tilde{\mathcal{D}} = \mathcal{D}_{\text{orig}}$
            \EndIf 
        \EndIf
        \State Observe environment state $o_t$ from $\tilde{\mathcal{D}}$
        \State Select action $a_t$ according to policy $\pi_\theta(a_t \mid o_t)$
        \State Compute $w_t = softmax(a_t)$
        \State Compute Oracle allocation $w^*(\tilde{\mathcal{D}}) = $ 
        \Statex \hspace{6em}  $ = \argmax_{w} \ \mathbf{Sharpe}(w, \bar \mu_t^{t+n}(\tilde{\mathcal{D}}), \bar \Sigma_{t-3n}^{t+3n}(\tilde{\mathcal{D}})) -$
        \Statex \hspace{8em} $ - TC_{\text{train}}(t) \cdot \|w - w_{t-1}\|_1 $

        \State Execute action $a_t$, transition to $o_{t+1}$ from $\tilde{\mathcal{D}}$
        \State Calculate Reward $r_t = -\bar \mu_t^{t+n}(\tilde{\mathcal{D}}) \cdot (w^* - w_t)'$
    \EndFor
    \State Compute advantage estimate $\hat{A}$ using GAE
    \State Add experience $(o_t, a_t, r_t, o_{t+1})$ to batch $D$
    
    \For{each RL training step}
        \State $\beta_{\text{entropy}}$ = $EntropySchedule(\text{step})$
        \State Recompute advantage estimate $\hat{A}$ using GAE
        \State Split batch $D$ into $K$ mini-batches $\mathcal{B}$
        \For{mini-batch $k = 1$ to $K$}
            \State Compute PPO total loss 
            \Statex \hspace{5em} $\mathcal{L} =\mathcal{L}_{\text{policy}} + \beta_{\text{entropy}} \cdot \mathcal{L}_{\text{entropy}} + \beta_{\text{value}} \cdot \mathcal{L}_{\text{value}}$
            \State Update critic and actor networks w.r.t $\mathcal{L}$
        \EndFor
    \EndFor
\EndFor
\State \Return $V_\phi, \pi_\theta$
\end{algorithmic}
\end{algorithm}

\section{Results and EMPIRICAL EVALUATION}
\subsection{Discussion}
We analyze the performance of the agent in Table \ref{tab:results1} using a variety of performance measures:

\begin{itemize}
    \item \textbf{Sortino Ratio}: A variation of the Sharpe ratio, it focuses on the standard deviation of negative asset returns to assess downside risk, offering a more targeted evaluation for investors.
    \item \textbf{Calmar Ratio}: This ratio measures downside risk by comparing the maximum drawdown to the average annual rate of return, providing insights into the risk-adjusted performance of an investment.
    \item \textbf{Omega Ratio}: Compares the probability of returns above a threshold to those below, considering the full return distribution.

\end{itemize}

Our PPO agent outperforms the 60/40 portfolio in terms of return across all periods, except for the validation set in the pre-pandemic phase. The Calmar ratio is significantly higher for regret PPO in most cases, with the exception of Phase 2. The Sharpe, Omega, and Sortino ratios are closely aligned. Notably, the validation performance for all measures in Phase 2 was suboptimal and shows room for improvement. The embedded drawdown approach has the lowest return among all approaches but typically achieves the best MDD, which aligns with the classic return-risk trade-off. Both the embedded drawdown and differential Sharpe reward functions are more conservative, consistently performing well in risk-accounting measures such as Sharpe, Sortino, and MDD. 
However, during the testing phase in the post-pandemic period, their performance was not as promising as the regret-based approach (-2.6\% for 60/40, -2.4\% for differential Sharpe, and -0.7\% for regret).

\begin{figure}[ht]
  \centering
  \begin{subfigure}{\linewidth}
    \centering
    \includegraphics[width=\linewidth]{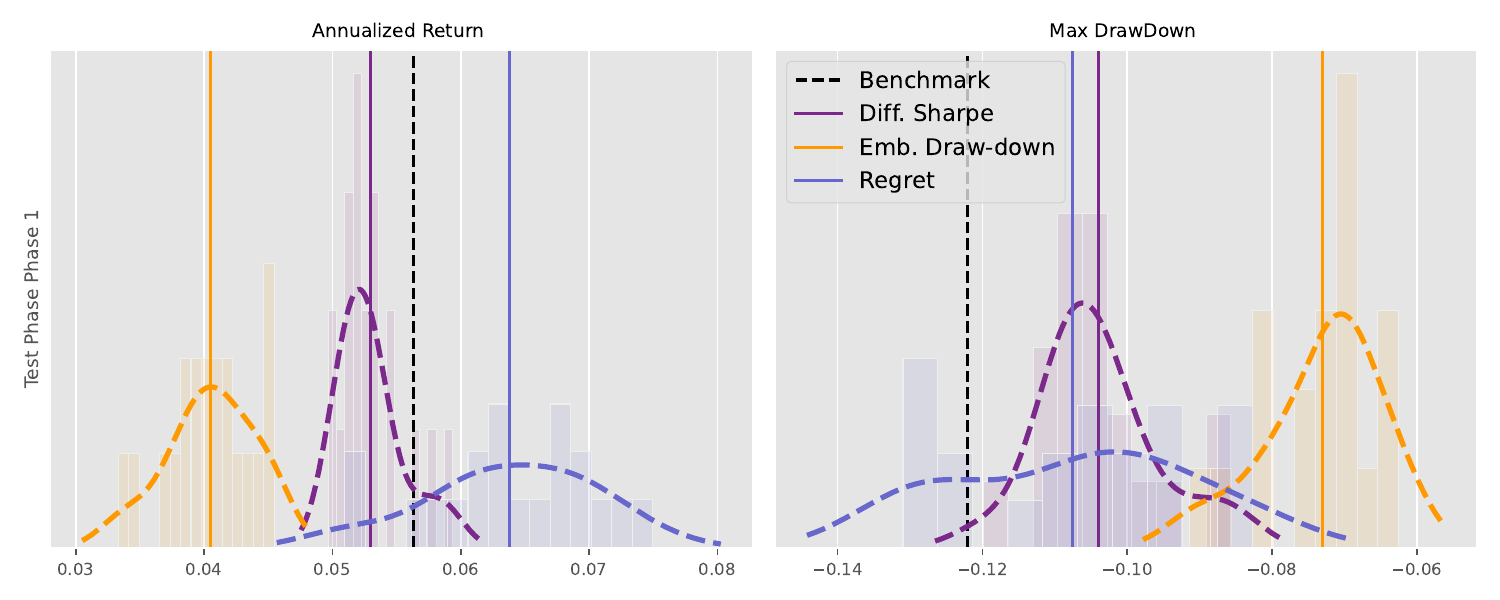}
    \label{fig:phase1}
  \end{subfigure}
  \begin{subfigure}{\linewidth}
    \centering
    \includegraphics[width=\linewidth]{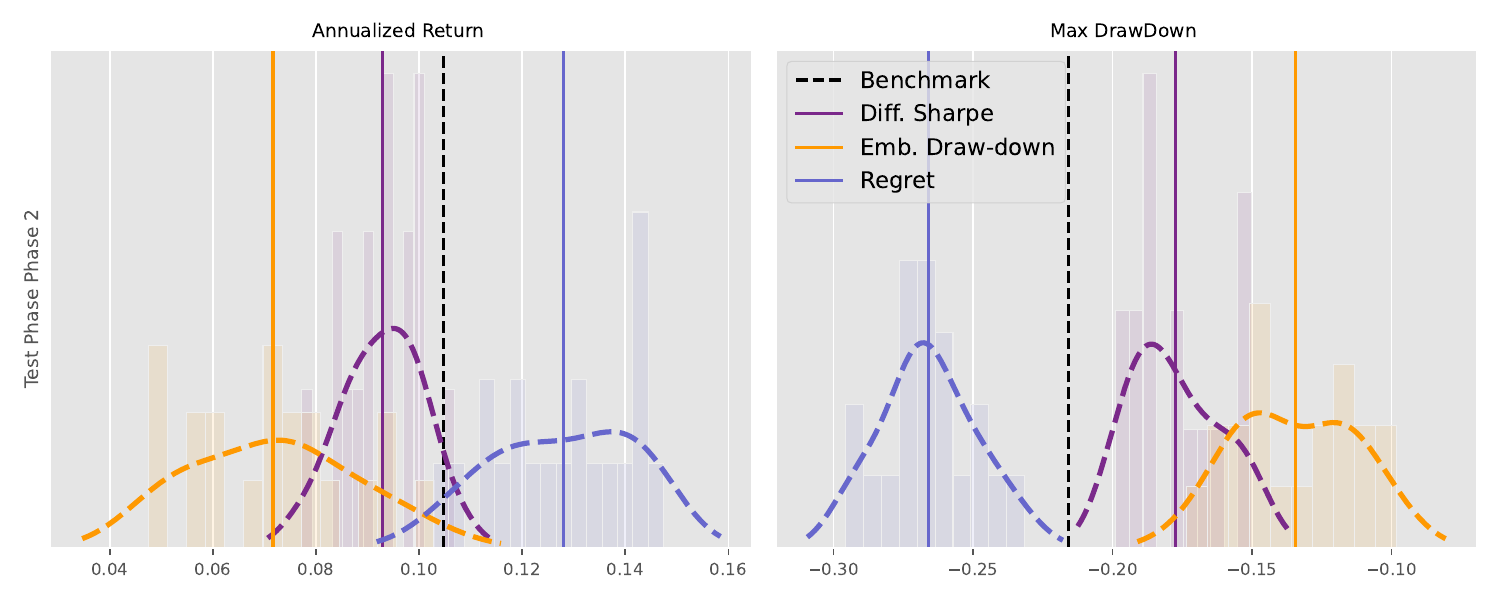}
    \label{fig:phase2}
  \end{subfigure}

  \begin{subfigure}{\linewidth}
    \centering
    \includegraphics[width=\linewidth]{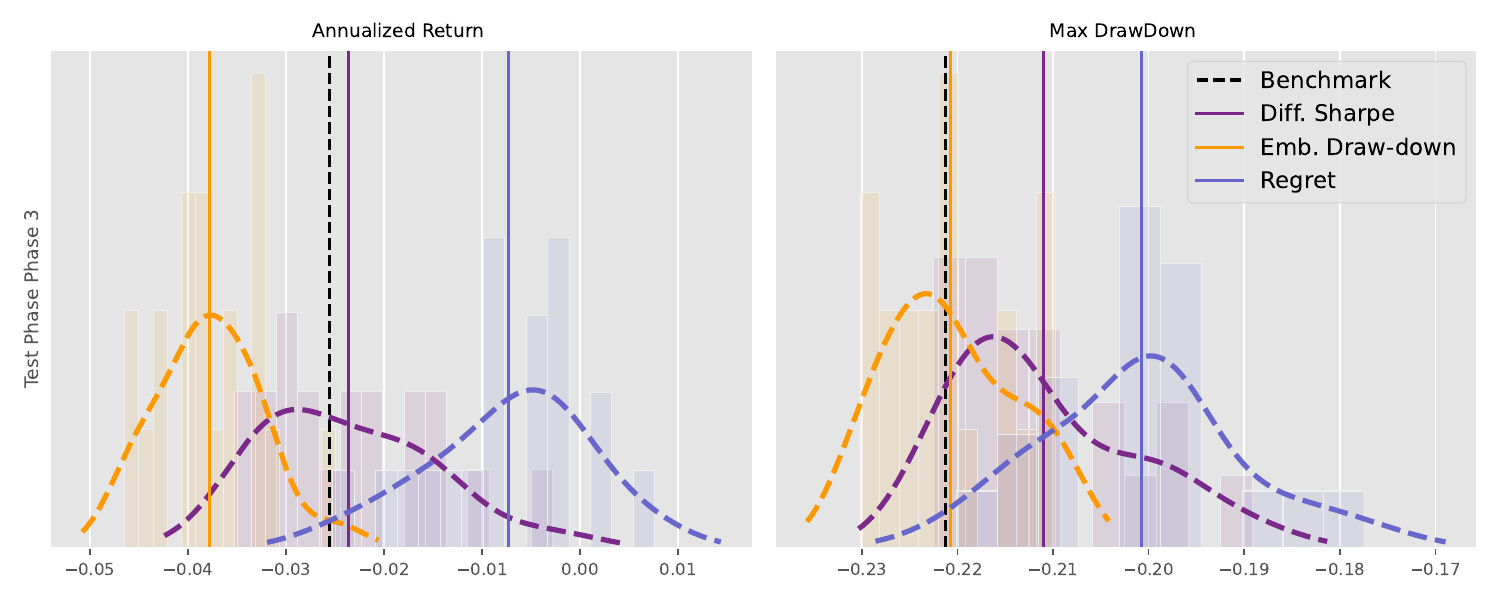}
    \label{fig:phase3}
  \end{subfigure}
  
  \caption{Distribution of returns and MDDs for the PPO agent for differential Sharpe, embedded drawdown and regret (ours) reward functions compared to the 60/40 benchmark (black dashed line) on the test periods. Our approach (in blue) focuses more on returns, allowing it to outperform all other methods in the primary objective across all three phases. Our model has proven superior to all considered approaches in the latest trading phase in both objectives.}
  \Description{Distribution of returns and MDDs for the PPO agent for differential Sharpe, embedded drawdown and regret (ours) reward functions compared to the 60/40 benchmark (black dashed line) on the test periods. Our approach (in blue) focuses more on returns, allowing it to outperform all other methods in the primary objective across all three phases. Our model has proven superior to all considered approaches in the latest trading phase in both objectives.}
  \label{fig:phases-plot}
\end{figure}

Overall, as illustrated in Figure \ref{fig:phases-plot}, our Sharpe regret-based approach with TC scheduling appears to be a preferable option, as it consistently outperforms the return benchmark and surpasses the MDD 60/40 benchmark in two out of three instances. Phase 1 is particularly challenging, with both measures showing flatter distributions. The transfer of weights in the later stages likely reduces variability in learning for the agent. Constant 60/40 only outperforms our approach in MDD during the pandemic breakout, which could be partly because the features are not optimal for detecting such signals. The MDD distribution in Phase 1 is more skewed compared to all other runs. As anticipated, the pandemic period (Phase 2) proved to be the most challenging for our agent, as its return distribution, while higher than the benchmark, remains relatively flat. Future work could focus on fine-tuning the agent to better identify and respond to crises through regime detection methods, as suggested in \cite{benhamou2021detecting, halperin_combining_2022}.


The embedded drawdown and differential Sharpe reward functions generate reasonable allocations but struggle to surpass the benchmark in the main measure: annual return. One possible reason is that these approaches do not explicitly account for transaction costs within the learning process. Agents using these reward functions tend to be more risk-averse, avoiding significant increases in positions in the riskiest assets. However, it is noteworthy that the other two approaches performed better in optimizing maximum drawdown compared to our regret-based approach. Interestingly, the embedded drawdown approach underperformed significantly during the post-pandemic period, barely beating the 60/40 benchmark. In the most recent testing period, our approach proved to be notably better than the other methods in both objectives.


  

Figures \ref{fig:Worcale} and \ref{fig:Alloc} demonstrate that regret-based agent generally aligns, though not as extremely, with the optimal allocation as shown by the allocations during the periods of April 2022, August 2022, and April 2023. Overall, the agent employing the Sharpe regret reward function tends to increase its position in risky assets during bullish markets and opts for a more conservative allocation when anticipating uncertain market conditions.

We also compare the effect of our modification on the training dynamics and the ability of the agent to generalize to the unseen data in Figure \ref{fig:novelt}. While the full configuration (TC + BB + Regret) is outperformed in the training environment, it shows a better and faster ability to generalize to the unseen data in the validation and test environments due to greater variability in the training set. Periodically introducing synthetic data, similar to many forms of regularization, increases the challenge for the agent to learn, as evidenced by the more fluctuating line in the training environment, but it helps the agent avoid memorizing non-reproducible effective allocation strategies. Interestingly, separately only BB and only TC configurations do not perform as well on the test set as their synergy. We can also observe that the current portfolio return (TC + BB + Return)  reward function is overfitting to the training data and is not managing to generate a profitable strategy out-of-sample.

\subsection{What Didn’t Work}

We also provide a brief discussion of the ideas that we tried and that did not work for our pipeline. Although these ideas failed for us, we hope that they might be helpful in other similar applications. DDPG and A2C algorithms failed to produce dynamic allocations and outputted nearly identical weights throughout the validation environment, regardless of the inputs in the observations. 

We also explored Convolutional Neural Networks (CNNs) for feature extraction, building on Benhamou et al. \cite{benhamou2021detecting}, who used multi-body policy networks with multiple tensors. We define the observation space with $k$ lags $L = [1, 2, 3, 6, 10, 15, 30, 60]$, yielding several tensors. The first network body utilizes asset information (\ref{eq:A_t}) as channels, while the second body processes market context (\ref{eq:C_t}) with 2-D convolution and a 3x3 filter. The approach underperformed compared to MLP architecture likely due to the transaction costs incurred from noisy allocation behaviors (Fig. \ref{fig:alloc_cnn}).

\begin{equation}
\footnotesize
A_t := \left[
\begin{array}{l}
\begin{bmatrix}
\mu_t \\
\mu_{t-l_1} \\
\vdots \\
\mu_{t-l_k}
\end{bmatrix} \in \mathbb{R}^{(k+1) \times 3 } , 
\begin{bmatrix}
\bar \mu_t \\
\bar \mu_{t-l_1} \\
\vdots \\
\bar \mu_{t-l_k}
\end{bmatrix} \in \mathbb{R}^{(k+1) \times 3 } ,
\begin{bmatrix}
\bar \sigma_t \\
\bar \sigma_{t-l_1} \\
\vdots \\
\bar \sigma_{t-l_k}
\end{bmatrix} \in \mathbb{R}^{(k+1) \times 3 }

\end{array}
\right] 
\label{eq:A_t}
\end{equation}

\begin{equation}
\footnotesize
C_t := \left[
\begin{array}{l}
\begin{bmatrix}
\alpha_t \\
\alpha_{t-l_1} \\
\vdots \\
\alpha_{t-l_k}
\end{bmatrix} \in \mathbb{R}^{(k+1) \times 3 } , 

\begin{bmatrix}
\bar q_t \\
\bar q_{t-l_1} \\
\vdots \\
\bar q_{t-l_k}
\end{bmatrix} \in \mathbb{R}^{(k+1) \times 3 }

\end{array}
\right] 
\label{eq:C_t}
\end{equation}

Alternatively, we also tried using differences in Sharpe ratios between optimal and taken allocations in the regret reward function. The agent with this reward function failed to learn a meaningful policy for unseen data, possibly due to the non-linear difference between optimal and taken actions. Additionally, we spent considerable time attempting to pre-train the agent with imitation learning using the Oracle as in \cite{kochliaridis2023combining} as an expert agent. All of the GAIL \cite{ho2016generative}, AIRL \cite{fu2018learningrobustrewardsadversarial}, and density-based reward modeling \cite{davis2011remarks} methods appeared to be highly sensitive to hyperparameters and did not result in improvement compared to training from scratch. Nevertheless, we see these approaches as promising for this application, and it may be our next direction for analysis.


\section{Future Work}
Our dynamic PPO regret-based agent consistently outperformed the return benchmark. The agent effectively adjusted allocations, increasing positions in bullish markets and adopting conservative allocations during uncertain periods. Our pipeline, which incorporates transaction cost scheduling, circular block bootstrap, and the future-looking regret-based reward function, demonstrates the ability to generalize to unseen test data. Discretizing the action space could enhance the model's scalability by reformulating the decision-making process. In such settings, ideas from \cite{elmachtoub2022smart} and \cite{mandi2020smart} may be explored for Oracle optimization.

Future work could involve further adapting the agent to operate in uncertain markets and respond more effectively during crisis events. Additional reward functions and their hyperparameters (such as Oracle's forecasting horizon) could be explored. The analysis of scalability may be conducted and the effect of larger portfolios and high-dimensional policies can be investigated.

We conducted a minimal, exploratory search for the following hyperparameters: the number of training episodes, early stopping time, the ramp-up duration for when the transaction cost (TC) scheduler levels off, and the convexity of the transaction cost. A more comprehensive exploration of these hyperparameters may be considered in future work.

Although we evaluated performance using MDD, it was not explicitly included in the reward, leading to less consistent performance in this measure. Expanding the observation space to a matrix by incorporating lagged variables could be beneficial; CNNs would be an appropriate choice for this setup. Additionally, dynamic strategies such as mean-reversion and momentum strategies could be integrated into the allocation process. More comprehensive hyperparameter tuning (on both the agent and environment sides) could be implemented. Given the sample inefficiency of RL approaches and computational constraints, we opted not to conduct an exhaustive hyperparameter search to avoid overfitting each validation set. Lastly, using narrower periods for the validation set could necessitate more frequent model recalibrations.








\bibliographystyle{ACM-Reference-Format} 


\appendix
\begingroup
\setlength{\floatsep}{5pt} 
\setlength{\textfloatsep}{5pt} 
\setlength{\intextsep}{5pt} 

\renewcommand{\arraystretch}{0.8} 

\captionsetup{skip=5pt} 



\pagebreak
\section*{Appendix}

\begin{table}[H]
    \footnotesize
    \caption{Hyperparameter values.}
    \begin{tabular}{l p{5cm} c}
        \toprule
        \textbf{Hyperparameter} & \textbf{Description} & \textbf{Value} \\ 
        \midrule
        $\gamma$ & Reward discount factor & 0.99 \\
        $lr$ & Learning rate & 0.001 \\
        $\epsilon$ & Range for clipping the gradients & 0.2 \\
        n\_steps & Number of steps for each update & 2048 \\ 
        batch\_size & Minibatch size & 64 \\ 
        $\beta_{\text{value}}$ & Value function coefficient for the loss calculation & 1 \\ 
        return\_lookback & Number of last business days to calculate rolling return & 40 \\ 
        std\_lookback & Number of last business days to calculate standard deviation & 60 \\ 
        n\_steps\_foresee ($n$) & Number of business days the Oracle can see into the future & 14 \\ 
        $\alpha$ & Convexity parameter for transaction costs & 1 and 0.45 \\ 
        $R_f$ & Risk-free rate & 0 \\ 
        $\beta_{\text{entropy}}$ (starting) & Starting entropy coefficient in the loss function & 0.00005 \\ 
        $TC_{\text{eval}}$ & Maximum transaction fees & 0.0025 \\ 
        Actor Network & Sizes of the (hidden) layers in the network & 19, 64, 64, 3 \\ 
        Critic Network & Sizes of the (hidden) layers in the network & 19, 64, 64, 1 \\ 
        \bottomrule
    \end{tabular}
    \label{tab:hp_values}
\end{table}

\begin{table}[H]

    \footnotesize
    \caption{Asset info.}
    \begin{tabular}{llccc}
        \toprule
        Name & Ticker Bloomberg  & \multicolumn{3}{c}{Strategy Weight} \\
        
        & & Strategy 1 & Strategy 2 & Strategy 3 \\
        \midrule
        Developed Equities & MXWOHEUR  & 1 & 0.55 & 0 \\
        Emerging equities & NDUEEGF   & 0 & 0.05 & 0 \\
        Global Credit & G0BC  & 0 & 0.2 & 0 \\
        Global Govies & W0G1  & 0 & 0.2 & 1 \\

        \bottomrule
    \end{tabular}
    \label{bloomberg_ticker}
\end{table}

\begin{table}[H]
  \footnotesize
  \caption{Timing phases with start and end dates for training, validation, and testing.}
  \centering
  \begin{tabular}{llccc} 
    \toprule
    &  & Phase 1 & Phase 2 & Phase 3 \\

    &  & \scriptsize{(pre-pandemic)} & \scriptsize{(pandemic)} & \scriptsize{(post-pandemic)} \\
    \midrule
    \multirow{2}{*}{Train} & Start Date & 1996-02-01 & 2002-01-01 & 2009-01-01 \\
                           & End Date & 2012-01-01  & 2016-01-01  & 2018-01-01  \\
    
    \multirow{2}{*}{Valid} & Start Date & 2012-01-01 & 2016-01-01 & 2018-01-01 \\
                           & End Date & 2015-01-01  & 2020-01-01  & 2022-01-01  \\
    
    \multirow{2}{*}{Test} & Start Date & 2015-01-01 & 2020-01-01 & 2022-01-01 \\
                          & End Date & 2020-01-01  & 2022-01-01  & 2024-01-01 \\
    \bottomrule
    
  \end{tabular}

  \label{tab:timing-phases}
\end{table}

\begin{figure}[H]
  \centering
  \includegraphics[width=0.8\linewidth]{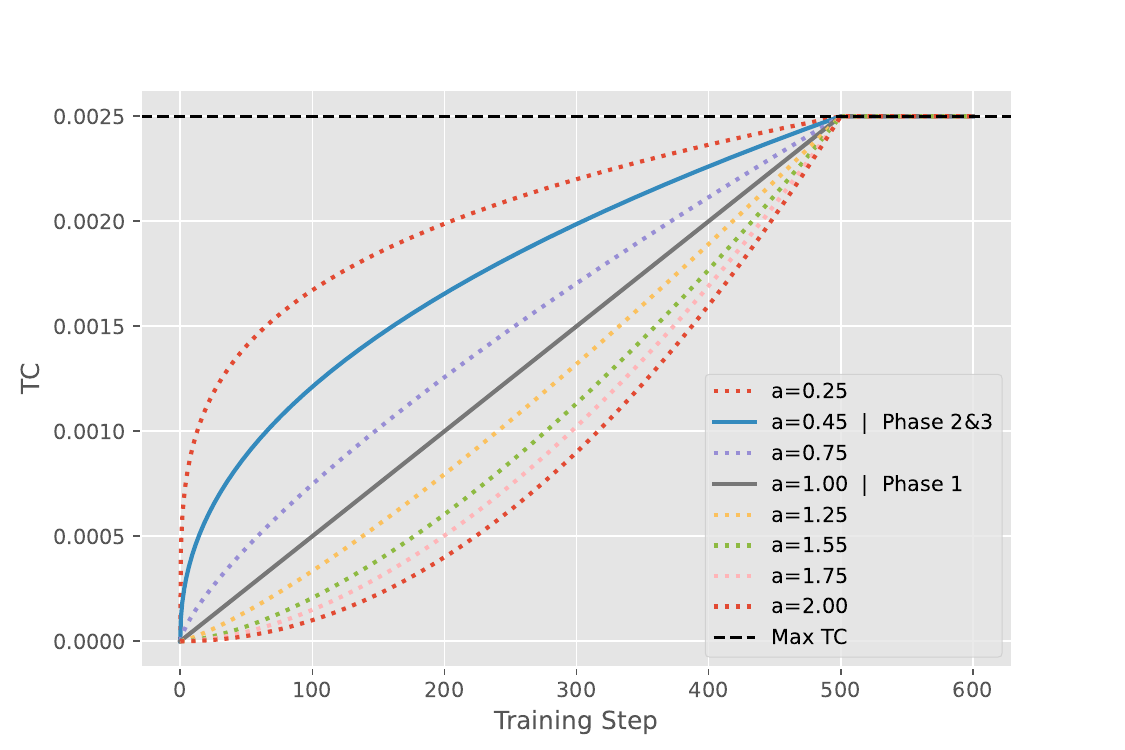}
  \caption{\small Transaction cost schedule example.}
  \label{fig:TC_sch}
  \Description{Transaction cost schedule example.}
\end{figure}

\begin{figure}[H]
  \centering
  \begin{subfigure}{\linewidth}
    \centering
  \includegraphics[width=\linewidth]{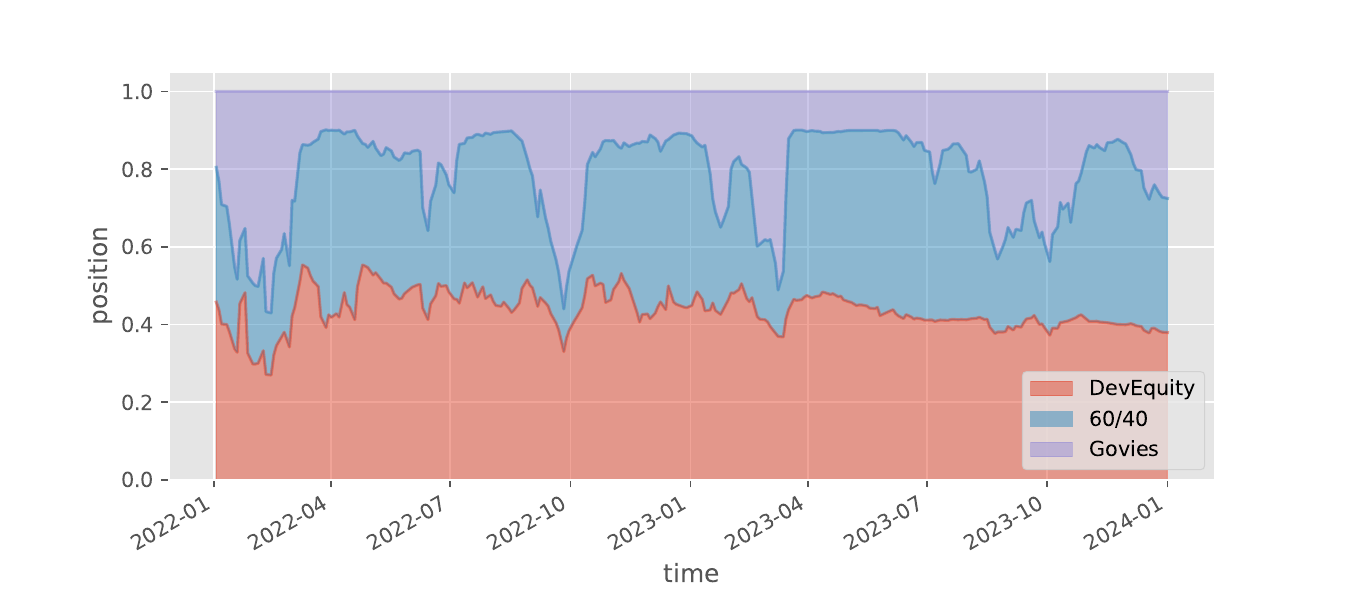}
  \label{fig:alloc_jpm}
  \end{subfigure}
  \begin{subfigure}{\linewidth}
    \centering
 \includegraphics[width=\linewidth]{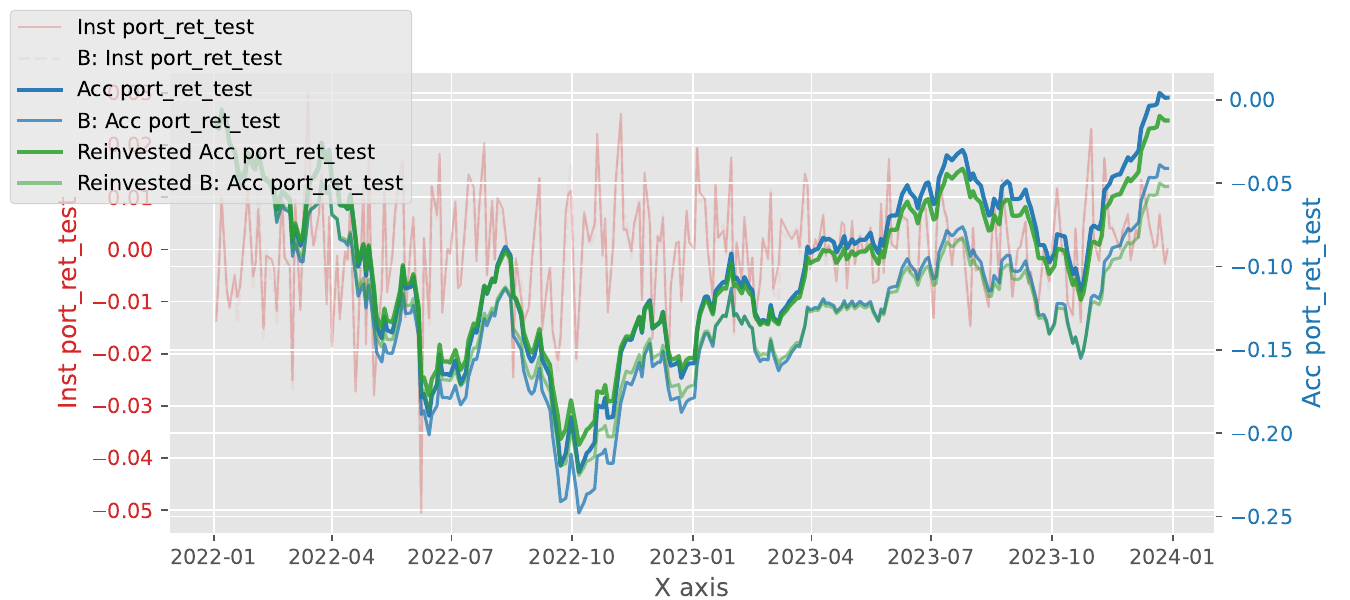}
  \label{fig:rets_jpm}
  \end{subfigure}

  \caption{ \small  Example of allocation and return dynamics by \\ Diff. Sharpe (\ref{eq:diff_sharpe}) during the testing period of phase 3.}
  \Description{  Example of allocation and return dynamics by \\ Diff. Sharpe (\ref{eq:diff_sharpe}) during the testing period of phase 3.}
  \label{fig:two-allocs1}
  
\end{figure}

\begin{figure}[H]
  \centering
  \begin{subfigure}{\linewidth}
    \centering
  \includegraphics[width=\linewidth]{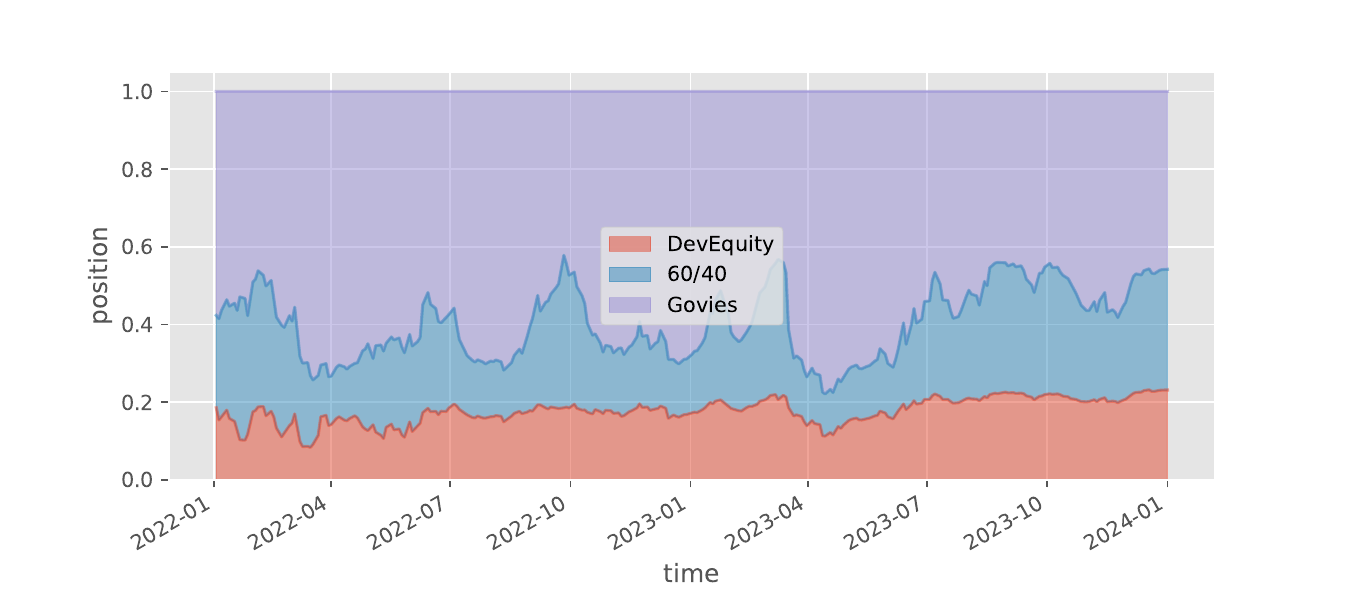}
  \label{fig:alloc_emb}
  \end{subfigure}
  \begin{subfigure}{\linewidth}
    \centering
 \includegraphics[width=\linewidth]{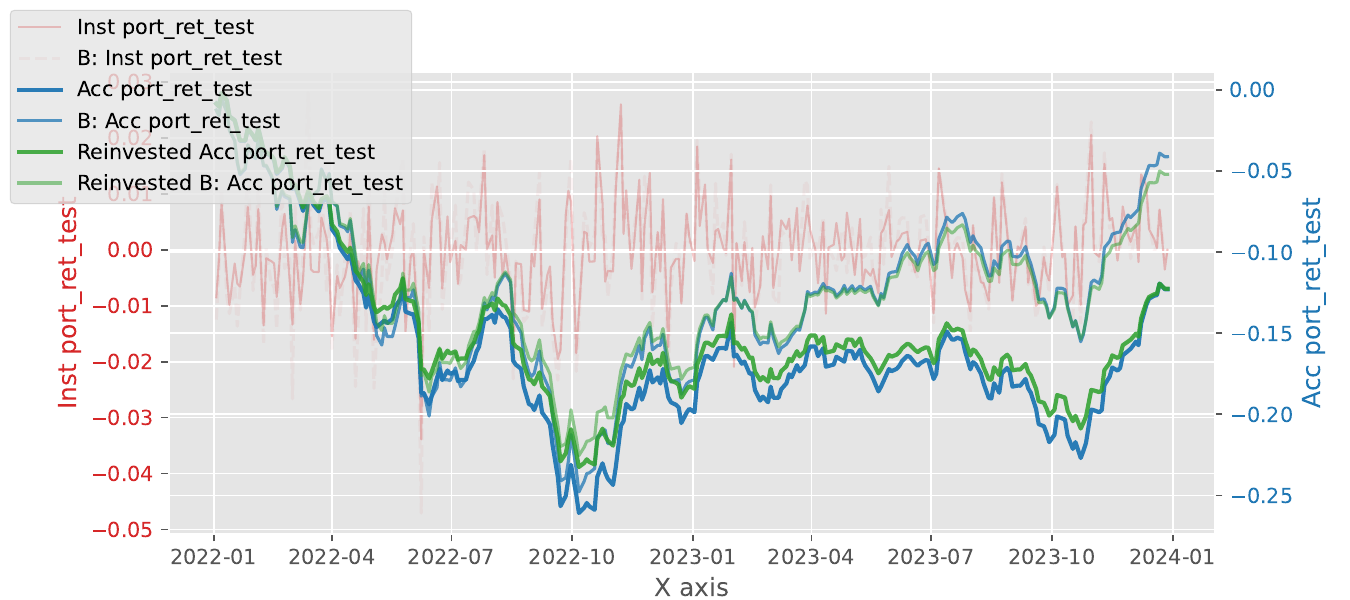}
  \label{fig:rets_emb}
  \end{subfigure}

  \caption{\small  Example of allocation and return dynamics by \\ Emb. Drawdown (\ref{eq:emb_rew}) during the testing period of phase 3.}
  \Description{ Example of allocation and return dynamics by \\ Emb. Drawdown (\ref{eq:emb_rew}) during the testing period of phase 3.}
  \label{fig:two-allocs2}
  
\end{figure}

\begin{figure}[H]
  \centering
  \includegraphics[width=\linewidth]{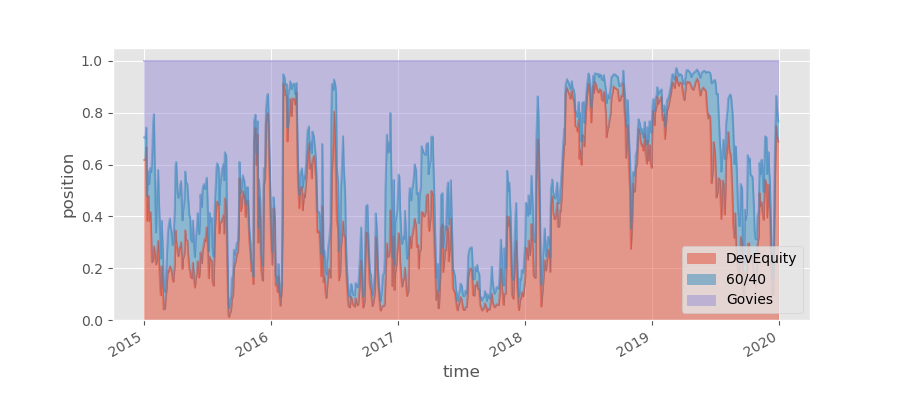}
  \caption{\small  Multi-Body CNN allocation example. We can observe noisy allocation behaviors. Introducing regularization to smooth the allocation line could enhance performance. For instance, adding a term to the reward function that penalizes sharp allocation changes—only to revert back—would indicate unnecessary transaction costs. Alternatively, we could include an integral of allocation changes as an additional penalty alongside our $\ell_1$ term.
  }
  \label{fig:alloc_cnn}
  \Description{ Multi-Body CNN allocation example. We can observe noisy allocation behaviors. Introducing regularization to smooth the allocation line could enhance performance. For instance, adding a term to the reward function that penalizes sharp allocation changes—only to revert back—would indicate unnecessary transaction costs. Alternatively, we could include an integral of allocation changes as an additional penalty alongside our $\ell_1$ term.}
\end{figure}

\section*{ \\ Glossary }
\label{sec:glossary}
\begin{scriptsize}
\begin{itemize}

    \item \textbf{Mean Return:} The expected return of an investment, denoted as:
    \[
    \mu = \mathbb{E}[R] = \begin{bmatrix}
        R_1 \\
        R_2 \\
        \vdots \\
        R_n
    \end{bmatrix}
    \]
    where \( R_i \) represents the financial return of asset \( i \).

    \item \textbf{Covariance Matrix:} A matrix that captures the variances and covariances of asset returns, represented as:
    \[
    \Sigma = 
    \begin{bmatrix}
        \sigma_1^2 & \sigma_{1,2} & \cdots & \sigma_{1,n} \\
        \sigma_{2,1} & \sigma_2^2 & \cdots & \sigma_{2,n} \\
        \vdots & \vdots & \ddots & \vdots \\
        \sigma_{n,1} & \sigma_{n,2} & \cdots & \sigma_n^2
    \end{bmatrix}
    \]
    where \( \sigma_i^2 \) is the variance of asset \( i \) and \( \sigma_{i,j} \) is the covariance between returns of assets \( i \) and \( j \).

    \item \textbf{Portfolio Return} ($\mu_p$): 
    \[
    \mu_p = w' \mu
    \]
    where $w$ is the vector of asset weights and $\mu$ is the return vector of individual assets.
  
\item \textbf{Portfolio Standard Deviation} ($\sigma_p$), \textit{Risk}:
    \[
    \sigma_p = \sqrt{w' \Sigma w}
    \]
    where $w$ is the asset weight vector and $\Sigma$ is the covariance matrix of asset returns.

    \item \textbf{Sharpe Ratio:} \label{eq:gl:sharpe} A measure of risk-adjusted return, calculated as:
    \[
    S = \frac{\mu_p - R_f}{\sigma_p}
    \]
    where \( \mu_p \) is the portfolio return, \( R_f \) is the risk-free rate, and \( \sigma_p \) is the standard deviation of the portfolio returns. The ratio represents return-risk trade-off.

    \item \textbf{Maximum Drawdown (MDD)}, \textit{Risk}: The maximum observed loss from a peak to a trough of a portfolio, calculated as:
    \[
    MDD = \max_{t} \left( \frac{P_{max} - P_t}{P_{max}} \right)
    \]
    where \( P_{max} \) is the peak portfolio value and \( P_t \) is the portfolio value at time \( t \).
\end{itemize}
\end{scriptsize}

\endgroup


\end{document}